\documentclass[aps, pre, notitlepage, twocolumn, showpacs, nofootinbib]{revtex4-1}

\usepackage{graphicx}
\usepackage{txfonts}
\usepackage{color}

\newcommand{\del}{\partial}

\begin{document}

\title{Buses of Cuernavaca - an agent-based model for universal random matrix behavior minimizing mutual information}

\author{Piotr Warcho\l{}} 
\email{piotr.warchol@uj.edu.pl} 
\affiliation{M. Smoluchowski Institute of Physics,  Jagiellonian University, PL--30--348 Cracow, Poland}

\date{\today}

\begin{abstract}

The public transportation system of Cuernavaca, Mexico, exhibits random matrix theory statistics. In particular, the fluctuation of times between the arrival of buses on a given bus stop, follows the Wigner surmise for the Gaussian Unitary Ensemble. To model this, we propose an agent-based approach in which each bus driver tries to optimize his arrival time to the next stop with respect to an estimated arrival time of his predecessor. We choose a particular form of the associated utility function and recover the appropriate distribution in numerical experiments for a certain value of the only parameter of the model. We then investigate whether this value of the parameter is otherwise distinguished within an information theoretic approach and give numerical evidence that indeed it is associated with a minimum of averaged pairwise mutual information. 

\end{abstract}



\maketitle

\section{Introduction}

As a galvanizing example of a socio-economic complex system exhibiting random matrix statistics, the buses of Cuernavaca \cite{KSbuses} are a poster child for universality \cite{RMTapli}. Thus, their behavior, like other phenomena of its kind, is not only interesting for its own sake, as a realization of some physical dynamical system, but it is also intriguing due to its connection to random matrix theory. Moreover, after $17$ years since the publishing of the seminal paper, the associated phenomenon is still being rediscovered in different systems like cars on a motorway \cite{cars} and subway trains in New York city \cite{nycmetro}.

As described by Krb\'alek and \v{S}eba, in Cuernavaca,  the buses do not have a given schedule. In contrast to most other cities, their drivers don't work for a single company. Instead, they operate according to their own interest, optimizing the amount of money earned while following a given route. Therefore, they strive to serve as many passengers as possible. This on the other hand means, that they tend not to arrive on a given bus stop immediately after their predecessors. To achieve that, they learn the information about the time of departure of the previous bus from collaborators stationed at the bus stop they arrive at. This mechanism seems to lead to the recreation of random matrix statistics for the unfolded \footnote{Through the unfolding procedure the mean distance between adjacent variable in the sample becomes $1$.} time intervals between arrivals of consecutive buses to a given bus stop, and the number variance (the variance of the number of buses arriving to a bus stop in a given time interval). 

Many models have been proposed to describe this phenomenon. In particular, the authors of \cite{BBDS}, famously formulate the problem as a set of independent, rate one, Poisson processes conditioned not to intersect. A cellular automaton and  a novel matrix model, the Damped Unitary Ensemble are proposed in \cite{KScella} and \cite{KHdue} respectively. In the first (\cite{BBDS}) case the relevant GUE statistics are recovered through an analytical calculation, yet the explanatory power of the model is diminished by the complicated mathematics involved. The two other models, on the other hand are dependent on a parameter, which drives the statistics from Poissonian to random matrix (GUE) like behavior. In the case of the cellular automaton, it is a ratio of the number of bus stops at which the bus drivers receive information and the number of all the bus stops. The size of the deterministic, off-diagonal elements of the Damped Unitary Ensemble matrices plays that role in the second case. We see the same transition in the aforementioned case of highway cars, with the traffic density as the driving factor, and for the New York city subway, where an effective value of Coulomb potential is assigned to the different behaviors. 

Inspired by an approach of combining Game Theory  and Random Matrix Theory to tackle performance optimization of multi-user telecommunication systems \cite{MIMO}, we introduce a novel model for simulating the bus system of Cuernavaca. In it, at each bus stop, each of the drivers optimizes his expected arrival time to the next stop with the use of a simple utility function. This function has one parameter which plays exactly the same role as the ones described above. Namely, it drives the statistics of the bus trajectories from Poissonian, through GUE to Gaussian. We then go one step further, and ask, whether the value of this parameter, which effectively measures the strength of the repulsive potential and for which we obtain the Random Matrix behavior, is otherwise singled out.  To answer this question, we turn to information theory. It turns out that this particular value is associated with a minimum of averaged mutual information between the trajectories of neighboring buses (Averaged Pairwise Mutual Information - APMI), measured at a given bus stop. We conjecture that it is this feature of the system that makes the Cuernavaca buses exhibit GUE statistics. 

This paper is organized as follows. We start by characterizing the proposed model and report the results of the simulation for the nearest neighbor spacing distribution and the number variance. This is followed by introducing the pairwise mutual information for this context and a presentation of the results of its estimation. We finish the paper with a brief conclusion and some conjectures. Finally, in appendix A, we show the simulation results for different numbers of buses and bus stops.

\section{The model}

First, let us describe our model for the bus system of Cuernavaca. Consider $N$ bus drivers traveling along a circular route with $K$ ordered stops. At each stop $k-1$, each of the drivers $n$, in turn learns their time of arrival $t_{n,k-1}$ and the time of arrival of their predecessor $t_{n-1,k-1}$. Based on that, and the knowledge of the route (namely the average time to travel between stops $k-1$ and $k$ denoted as $\tau_{k}$ \footnote{For cyclic boundary conditions -  the route forming a loop - $\tau_{1}$ is the average travel time between stops $K$ and 1} ), they assess at what time $t_{n,k} $ to arrive at the next bus stop to maximize their earnings. Thus, each maximizes some utility function $u_{n,k}$ for the arrival of the $n$'th bus to the $k$'th stop, with respect to the intended arrival time $t_{n,k}$. 
The amount of people entering the bus and the arrival time itself (due to unexpected occurrences on the route) will in the end be random variables. We will however neglect that and work with the utility function not its expectation. Thus, in the end, at each bus stop, each bus driver solves
\begin{eqnarray}
\frac{\del u_{n,k}}{\del t_{n,k}}=0
\end{eqnarray}
to find the optimal arrival time.

Now, let us design the utility function. We know that the bus drivers will strongly avoid not taking anybody at the stop, which happens when the time distances $(t_{n,k}-t_{n-1,k})$ between arrivals are very short. On the other hand, if they take a sufficiently large number of people, they won't care as much about taking a few more. This (as well as Random Matrix Theory) suggests taking the logarithmic  function to represent such behavior. It will be the repulsive core of the potential. The difference between the arrival times will not be expressed in terms of $t_{n-1,k}$ as this is not known for the $n$'th driver yet (at bus stop $k-1$). Instead, it is expressed with the help of the average travel time between the stops and the known arrival time of bus $n-1$ to the stop $k-1$ as $t_{n,k}-t_{n-1,k-1}-\tau_k$.
The journey, moreover, cannot be too long and the passengers value how predictable is the time it is going to take. We will use the harmonic potential to represent the fact that the travel time should be close to the average. Thus, we set:
\begin{eqnarray}
u_{n,k}=b\, {\rm ln}\left(t_{n,k}-t_{n-1,k-1} - \tau_k\right)-\left(t_{n,k}-t_{n,k-1}-\tau_{k}\right)^2
\end{eqnarray}
and therefore each of the drivers, solves the following quadratic equation
\begin{eqnarray}
b=2\left(t_{n,k}-t_{n,k-1}-\tau_{k}\right)\left({t_{n,k}-t_{n-1,k-1}-\tau_k}\right)
\end{eqnarray}
and chooses the solution consistent with $t_{n,k}>t_{n,k-1}$. This can be seen as $N$, mutually dependent nonlinear maps. $b$ is some yet unknown parameter, which we assume to be the same for all the bus drivers in the model and which marks the strength of the logarithmic repulsion with respect to the harmonic potential. The presented approach is convenient as we can explicitly solve the quadratic equation and therefore, as we describe in the next section, simulating this model on a computer is straightforward.

Thus, there are many agents (bus drivers), each optimizing the behavior with respect to the agent in front. Moreover, a change in the arrival time with respect to the optimal one, for some bus driver, will by definition result on average with a smaller pay-off.  The realization of this scheme is thus a form of Nash equilibrium \cite{NetworkinGames} - each player, at every bus stop can choose from a continuous set of strategies associated with the travel times. Note, that this does not generalize to a Wardrop equilibrium \cite{NetworkinGames, Wardrop} as the number of drivers grow, because the choice of the strategy of one, directly influences the situation of only the one immediately behind him. Finally, the optimization is done individually and for each bus stop separately, hence there might be a global equilibrium possible, for which the total earnings of all the bus drivers are higher then the total earning resulting with using this optimization method - this being a hallmark of decentralized systems. A centralized system, in other words, would (possibly) be able to decide on arrival times to the proceeding stops of each of the buses in such a way to achieve a higher sum of earning of all the bus drivers. In this case, the Nash equilibrium would not be optimal (at least globally). This is akin to the Braess's paradox describing situations for which the overall performance of a traffic network is not optimal when the drivers choose their route selfishly.

\section{Results of the simulation}

Now, our goal is to compare this model to the real bus system of Cuernavaca. To this end, for the simulated system we will compute the two statistical properties studied in the original paper by Krb\'alek and \v{S}eba - the unfolded nearest neighbor spacing distribution and the so called number variance. Under the condition that, on average, the time distances between the buses arriving at a given stop are one (realized by the unfolding procedure), the former is the distribution ($P(s)$) of time distances ($s$) between the arrivals of the neighboring buses. The latter ($N(t)$), under the same condition, measures the variance of the number of buses arriving in a time interval of length $t$. For random matrices of the Gaussian Unitary Ensemble, these are given by the Wigner surmise approximation:
\begin{eqnarray}
P(s)=\frac{32}{\pi^2}s^2 e^{-\frac 4\pi s^2}
\end{eqnarray}
and 
\begin{eqnarray}
N(t)\approx \frac{1}{\pi^2}\left({\rm log}2\pi t +\gamma +1\right)
\end{eqnarray}
respectively \cite{Mehta}, and where $\gamma$ is the Euler constant.

\begin{figure}[htbp]
	\includegraphics[width=0.42\textwidth]{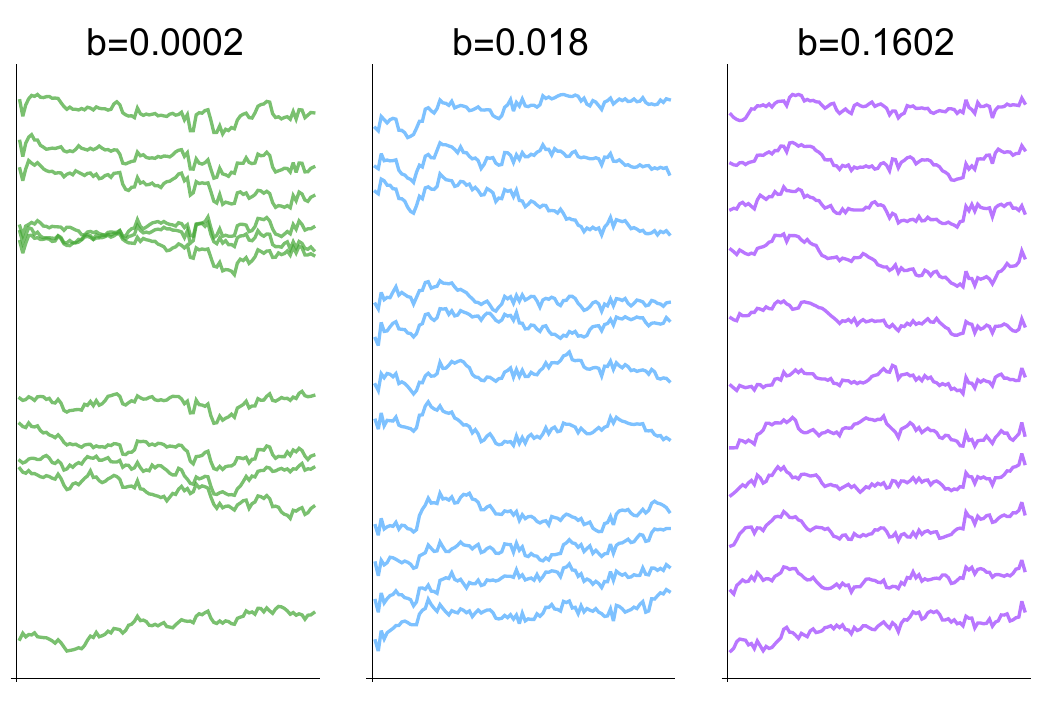}
	\caption{Eleven sample bus trajectories, after unfolding, for three different values of $b$. The horizontal axes represents the consecutive bus stops, and the vertical one, the unfolded arrival times.}
	\label{fig:tr}
\end{figure}

Thus, for our simulation, we generate $100$ buses traveling around a cyclic root with  $100$ stops distributed so that the travel time between the adjacent stops ($\tau_k$) is given by a uniform distribution, here on the interval $[1,1.5]$. The latter sets the  time scale of the experiment. We initialize the simulation by setting the times of arrival of the first bus (labeled with $1$) to all the stops on the first cycle.  These are normally distributed, with mean $t_{1,k-1}+\tau_k$ and variance $0.05$, which constitutes $4\%$ of the average $\tau_k$ across the bus stops.  $t_{1,1}$ is set to be $0$. Then, we set the times of arrival to the first stop in the first cycle for each of the rest of the bus drivers. These start their journeys in equal time intervals such that on average, the last leaves the first stop when the first driver arrives (on average) at the last one. The rest of the simulation is performed through the optimization method described above and with a cyclic boundary condition, so that when the bus driver leaves bus stop $K$ it goes towards the one labeled by $1$. The arrival time is at each step perturbed, with respect to the one resulting from the optimization procedure, by a random number from the normal distribution with mean $0$ and variance $\sigma=0.05$ (again, $4\%$ of the average $\tau_k$ across the bus stops). Here, we keep $\sigma$ fixed, however this is a parameter that has some amount of qualitative influence on the results - we will come back to this later. For now we just mention that it has to be small, so that bus overtakings due to noise are extremely rare. 

\begin{figure}[htbp]
   \includegraphics[width=0.49\textwidth]{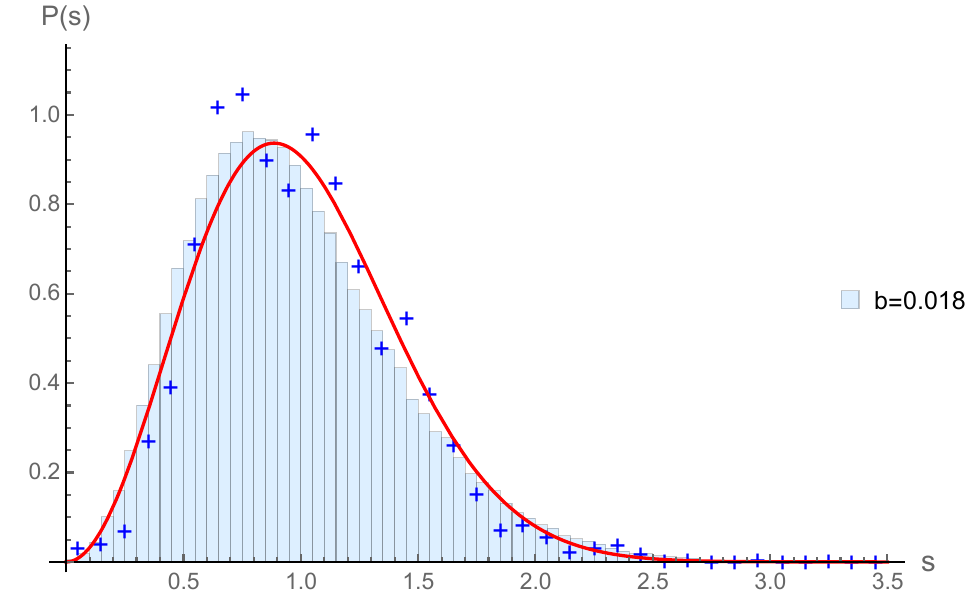}
	\caption{Nearest neighbor spacing distribution. The red line is the the analytic approximation (the so-called Wigner surmise) of the GUE random matrix result. The blue crosses are results obtained in \cite{KSbuses} for the Cuernavaca buses whereas the blue histogram is the results of the numerical simulations for $b=0.018$.  }
	\label{fig:es1}
\end{figure}

\begin{figure}[htbp]
   \includegraphics[width=0.49\textwidth]{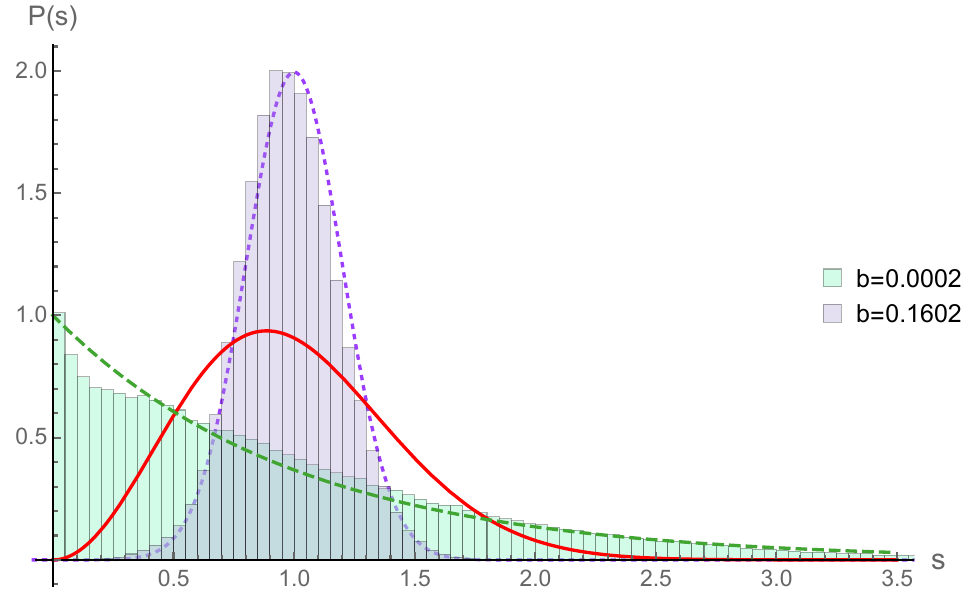}
	\caption{Nearest neighbor spacing distribution. The red line is the the analytic approximation (the so-called Wigner surmise) of the GUE random matrix result. The histograms are the results of the numerical simulations for $b=0.0002$ (green) and $b=0.1602$ (purple). The dashed green and purple lines mark the exponential and normal (best fit, with mean $1$) distributions respectively. }
	\label{fig:es2}
\end{figure}

Altogether, we perform $100$ such experiments, with $32$ cycles, for each given value of $b$ (between $0.0002$ and $0.1602$ with a step of $0.002$). The distribution of arrival times for a given stop in a given cycle is already, on average, uniform. Thus, what the unfolding procedure effectively does, it  rescales the arrival times so that they are all, on average, uniformly distributed on an interval from $0$ to $N$.

We showcase some of the unfolded bus trajectories generated with the numerical simulations in figure \ref{fig:tr}.
For large values of $b$, up to some small fluctuations, the buses travel in unison, whereas for small ones, there is little coordination on  larger timescales. The associated nearest neighbor distance distributions can be seen in figures \ref{fig:es1} and \ref{fig:es2}. The former plot depicts it for the value of $b$ for which the ($\beta=2$) Wigner surmise and the histogram match best. The skewness of the distribution obtained with the model is the result of the value of $\sigma$ being small but finite. Our experiments showed, that making $\sigma$ smaller, reduces the skewness to the point of a very good match with the random matrix result  \footnote{We decided to keep $\sigma$ at $4\%$ relative to the average time distance between the buses, as this seems realistic, and because we wanted to showcase the skewness effect.}. A consequence of this modification is a rescaling of $b$ (the essence of the information measure result shown below does not change however). This is because if random occurrences on the route have more influence on the arrival time, this has to be balanced by the strength of the logarithmic potential to  generate random matrix statistics for the nearest neighbor distribution. As can bee seen in the second plot, for significantly smaller $b$'s, the distribution approaches the Poissonian statistics, as the harmonic potential dominates. For large values of $b$ it becomes Gaussian. Then, the logarithmic part of the 'potential' has the upper hand and this behavior comes from  the random (normal) perturbation of the arrival times with respect to the optimal arrival time the bus drivers strived to achieve.  

\begin{figure}[htbp]
	\includegraphics[width=0.42\textwidth]{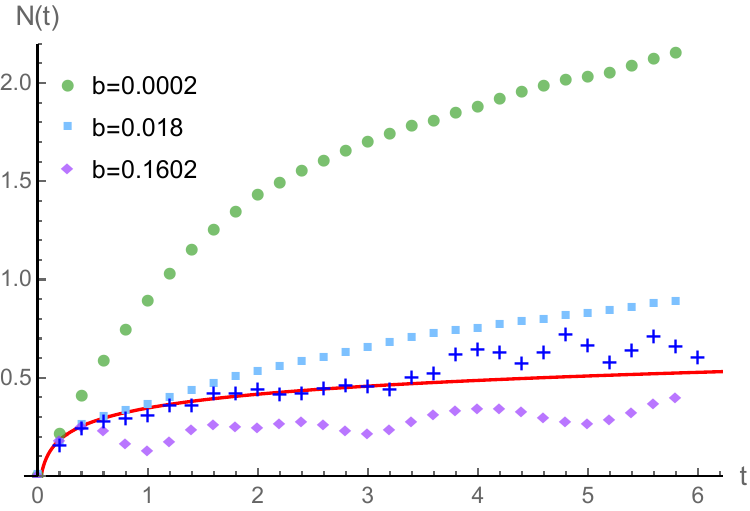}
	\caption{The number variance and its dependence on the time interval in consideration. The red line is the analytic, GUE result. The blue crosses are results obtained in \cite{KSbuses} for the Cuernavaca buses. The green disks, the blue squares and the purple diamonds are the results for the numerical simulations for $b$ equal $0.0002$, $0.018$ and $0.1602$. }
	\label{fig:nv1}
\end{figure}

The equivalent of the random matrix theory number variance is the variance of the distribution of the number of buses arriving at a certain bus stop in a given time interval. This is looked at as a function of the length of that time interval. The result of the simulation are depicted in figure \ref{fig:nv1} for the value of $b$ associated with GUE statistics and the two values of $b$ on the border of the $b$ interval of interest respectively. Note that one needs quite a lot of data to obtain a decent estimation of the number variance behavior. Here each point of the plot is result of an average of $51600$ measurements in the simulations. Considering the results of the Cuernavaca measurements were based on a smaller sample, we may say that the behavior of our model for $b=0.018$ is reasonably consistent with reality.
\section{Analysis in terms of Mutual Information}

Now that we know that our model, to an extent, reassembles the behavior of the bus system in Cuernavaca for a certain value of parameter $b$, we want to ask why is this particular value so special. (As $b$'s counterpart in Random matrix Theory is $\beta$ and for the bus system $\beta$ manifests as being equal to $2$, this touches a more general question of the Universality of the Gaussian Unitary Ensemble). In other words, ideally we would like to know why the Cuernavaca buses exhibit random matrix, GUE statistics. We will not be able to answer this question here, however we conjecture, that it can be rephrased in terms of information theory. Namely, provided there exists an entropy based measure, that calculated for the process described by our model has an extremum for the particular $b$ associated with the Cuernavaca buses, we can instead ask why does the associated extremization principle work. In turns out such an entropic measure exists - we have looked at entropy, multivariate entropy, pairwise \cite{SHte, JBCtecrit} and global entropy transfer, as well as global information \cite{Ising}, and it turns out that this measure is the pairwise mutual information. Let us now formally introduce this last quantity.

Let $W_n$ be the arrival time (after unfolding) of a given bus $n$ to some bus stop - we treat it as a random variable. The associated differential entropy of the probability distribution of $W_n$ reads: 
\begin{eqnarray}
H(W_n; b)=-\int_{-\infty}^{+\infty} p(w_n; b)\,{\rm ln} p(w_n ; b)\,{\rm d}w_n
\end{eqnarray}
For a given bus $n$, each (unfolded) arrival time $w_n$ (to any of the bus stops) is now a realization of the, assumed to be stationary, stochastic process resulting from the evolution given by our model. 
The joint differential entropy of two such random variables is:
\begin{eqnarray} \nonumber
H_j(W_i, W_j; b)= \\
-\int_{-\infty}^{+\infty} \int_{-\infty}^{+\infty} p(w_i, w_j ; b)\,{\rm ln} p(w_i, w_j; b)\,{\rm d}w_i {\rm d}w_j
\end{eqnarray}
and measures the uncertainty associated with some two bus trajectories. Importantly, here we are interested in the case where $W_i$ and $W_j$ represent the arrival times to the same bus stop.
Thus, the joint random variable is the tuple $(W_i,\,W_j)$ associated with a given bus stop and the associated joint probability density can be estimated based on its realizations for different bus stops.
Now, the pairwise mutual information of the neighboring buses, in terms of entropy is:
\begin{eqnarray} \nonumber
I(W_n :W_{n+1} ; b)=\\
H(W_n; b)+H(W_{n+1}; b)- H(W_n, W_{n+1}; b).
\end{eqnarray}
Finally, the averaging is done over all the vehicles.
Therefore, APMI measures the reduction of uncertainty in the arrival time (subject to the unfolding procedure done across the arrival times to a given bus stop) of one of the two neighboring buses which stems from the knowledge of the arrival time (to the same bus stop) of the other and vice-versa. In other words, it is the information shared by these two random variables.

To calculate it, we use the unfolded arrival times of the last $29$ (of the generated $32$, to allow for 'thermalization') cycles of $100$ stops, for each nearest neighbor pairs of the $100$ buses and perform this experiment $100$ times. The Kraskov-Stoegbauer-Grassberger estimation method \cite{Kraskov}, as implemented in the JIDT package \cite{JIDT} as Kraskov algorithm $1$ is utilized.

\begin{figure}[htbp] 
	\includegraphics[width=0.45\textwidth]{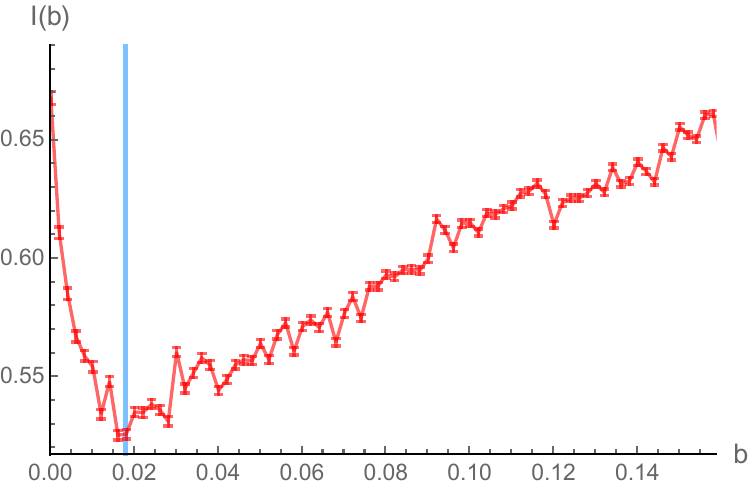}
	\caption{In red, the averaged pairwise mutual information (measured in nats) for values of $b$ in $0.002$ intervals between $0.0002$ and $0.1602$. We plot the error bars representing the standard error of the mean of the results from all the experiments. The vertical blue line marks $b=0.018$.}
	\label{fig:pmi}
\end{figure}

We can see the results in figure \ref{fig:pmi}. The pairwise mutual information, within the studied spectrum of $b$, reaches its minimum close to the value of $b$ related to the GUE. Thus, as advertised earlier, we claim that the bus drivers of Cuernavaca, while maximizing their payoff, they minimize the mutual information with respect to the neighboring buses. This in turn leads to the manifestation of the GUE statistics of the nearest neighbor spacing distribution.

Let us note that it is clear why APMI is large for large values of $b$ - the bus trajectories are rigid with respect to each other and the amount of shared information is high. Why is the value large for very small $b$'s? This is because then, it often happens that the buses drive close to each other so that their mutual rate of influence through the logarithmic potential increases. Moreover the overall entropy of the trajectories is large for small $b$'s. In fact the relative APMI (APMI divided by the averaged entropy of the particular trajectories, namely the fraction of shared information and the uncertainty) is monotonic across the spectrum of $b$, rising with growing $b$. This is in accordance with the intuition, that when the trajectories are more and more random, the relative amount of information they share about each other decreases. This is captured by Mutual Information relative to the overall entropy.

In appendix A we extend this numerical analysis to different values of $N$ and $K$. In particular we show how the value of $b$ associated with the minimum o APMI, converges to the $b$ related to GUE statistics in growing $N$(=K) and how the results change for $N\ne K$.

\section{Summary and Conclusions}

In this paper we have proposed a straightforward, agent-based approach for modeling the bus system of Cuernavaca. The resulting nearest neighbor distance distribution and number variance compare favorably to the associated results for random matrices and the bus system itself.  The proposed model has one adjustable parameter, however we showed that choosing its value is linked to a minimization of averaged pairwise mutual information. This enables us to link the global statistical behavior of this complex system to, both, a maximization of some profit function of individual drivers and a minimization of the information shared between the nearest neighbors. Note however, that the mechanism behind the latter is still unclear. Perhaps there is some optimization process not taken into account, otherwise one is tempted to see the minimization of the APMI as a version of a kind of Maximum Entropy Principle at play for complex interacting systems. Either way, it would be very interesting if the same effect is visible in the parameter dependent models mentioned in the introduction.

Finally, one can generalize the so-called classical Gaussian random matrix ensembles of $\beta=1,2,4$ to arbitrary real, positive $\beta$ \cite{DE}. These in turn have their own nearest neighbor spacing distributions starting form the Poisson distribution for $\beta=0$ \cite{CMD}. We conjecture, that the studied pairwise mutual information will also have a minimum for stochastic processes generating arbitrary-$\beta$ ensembles, as a function of $\beta$, for $\beta=2$. Perhaps the constraint of Gaussianity can also be alleviated and the last statement is true for matrix ensembles with random elements generated from a range of distributions, namely the General $\beta$-Ensembles \cite{BEY}. If this is the case, it may have important consequences for systems which exhibit local random matrix and semi-Poisson statistics and may help with understanding of the universality properties of the Gaussian Unitary Ensemble. We plan to verify this with numerical experiments.

\begin{acknowledgments}
We would like to acknowledge the funding of the Polish National Science Centre through the projects DEC-2011/02/A/ST1/00119 and  2016/21/D/ST2/01142.
Various discussions on the subject with Maciej A. Nowak, Jacek Grela and Jeremi Ochab are greatly appreciated.
\end{acknowledgments}

\appendix

\section{APMI for different $N$ and $K$}

Here, we will study the APMI for different numbers of buses and bus stops. The simulation are done in the same way as in the main part of the paper, except now we cover the spectrum of $b$ between $0.0002$ and $0.0782$ in intervals of $0.002$. When $N$ and $K$ are varied, for each value of $b$, the number of experiments is the integer part of $5000/N$ and the number of cycles for a bus to perform in any experiment is the integer part of $2500/K$. This way the statistic samples are of roughly the same size. For $N=K=100$, the setup from the main part of the paper is used.

\begin{figure}[htbp] 
	\includegraphics[width=0.45\textwidth]{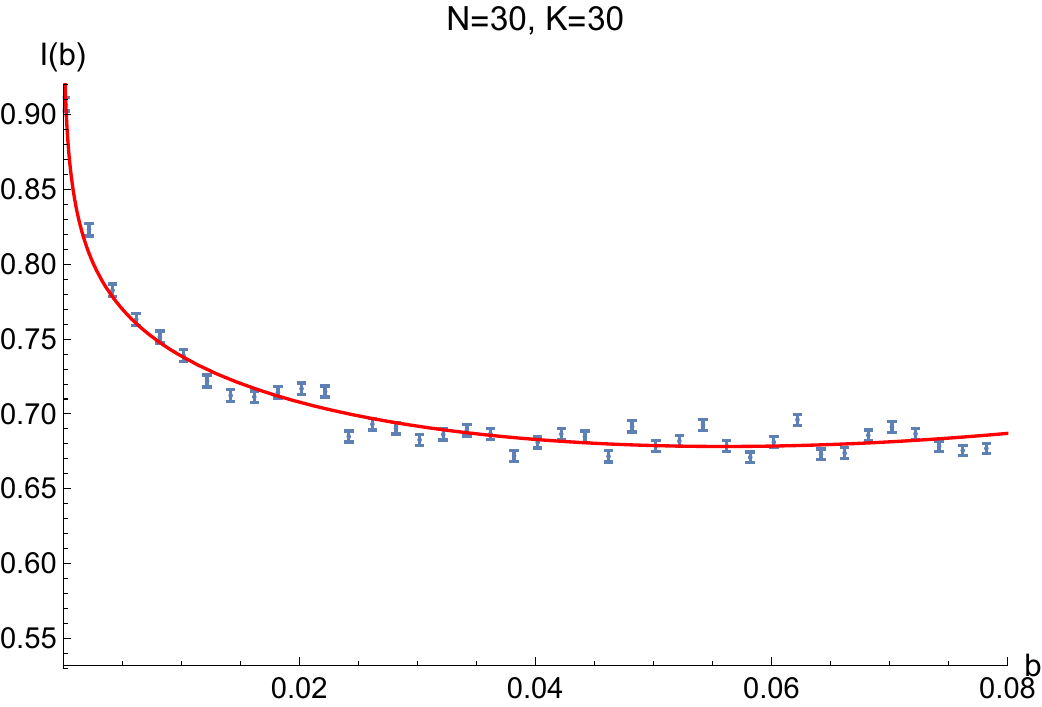}
	\includegraphics[width=0.45\textwidth]{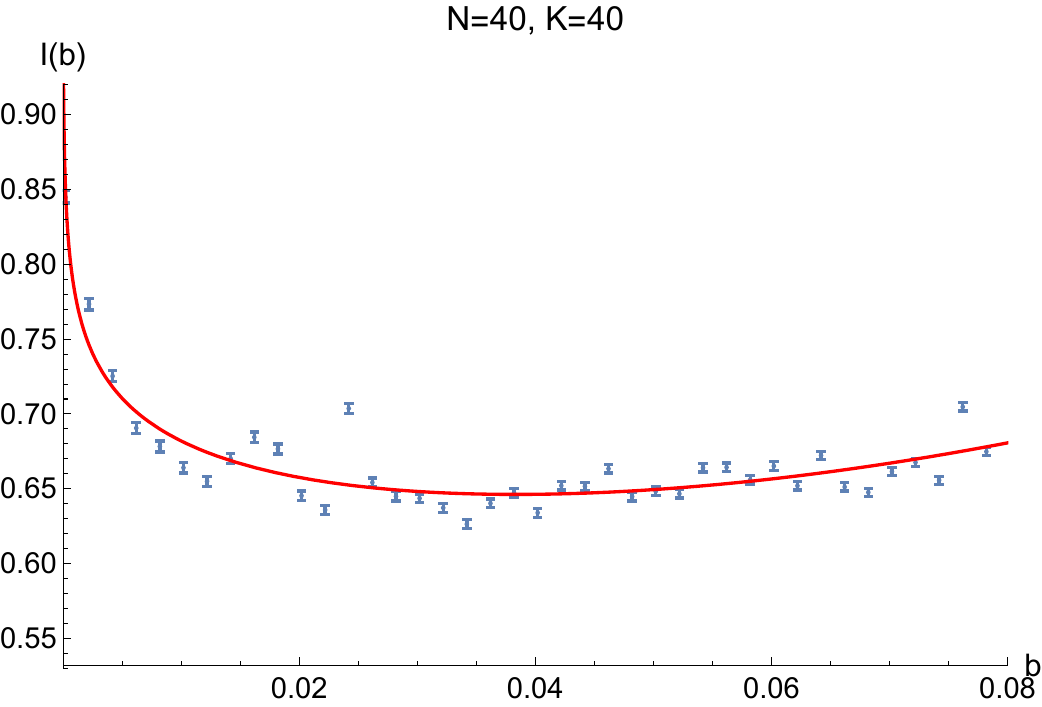}

	\includegraphics[width=0.45\textwidth]{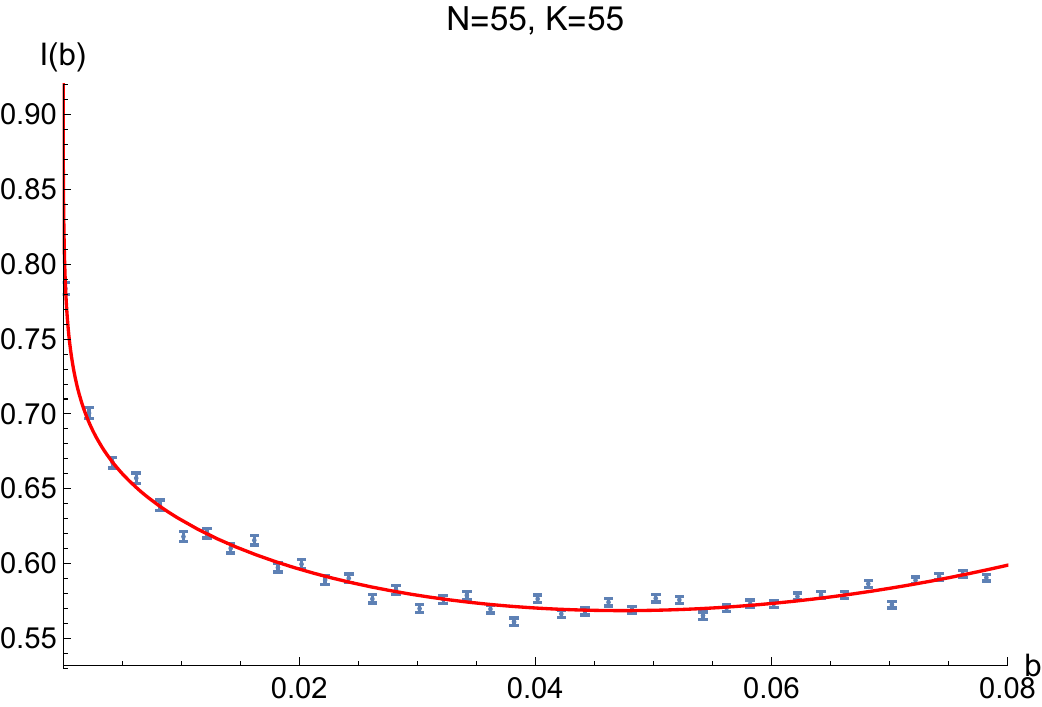}
	\includegraphics[width=0.45\textwidth]{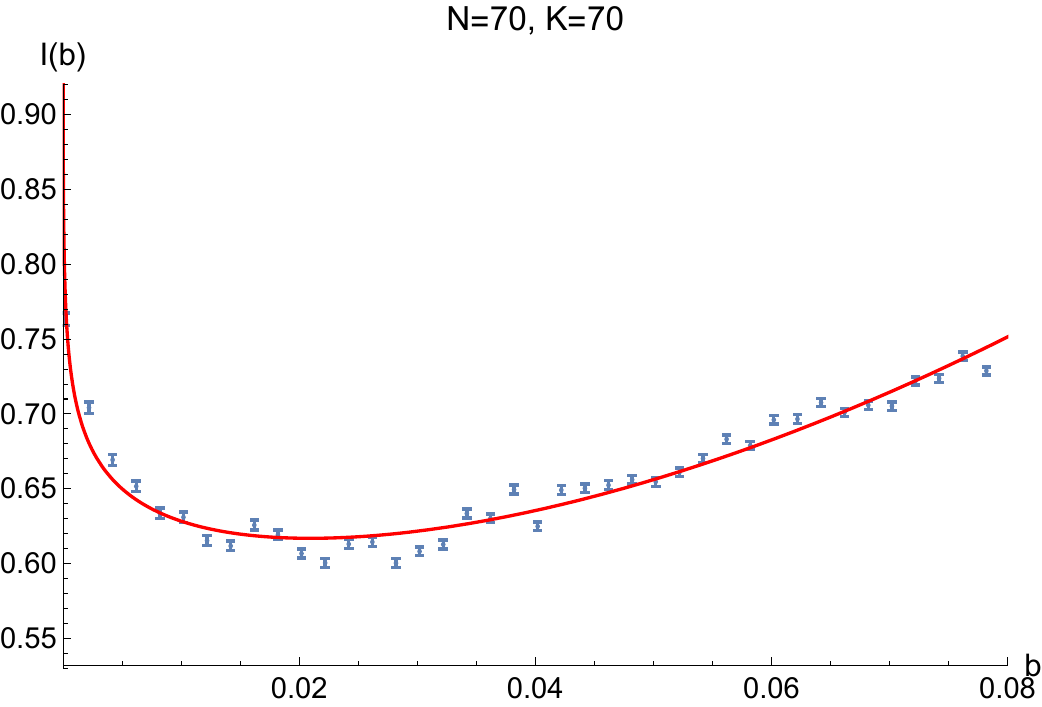}

	\includegraphics[width=0.45\textwidth]{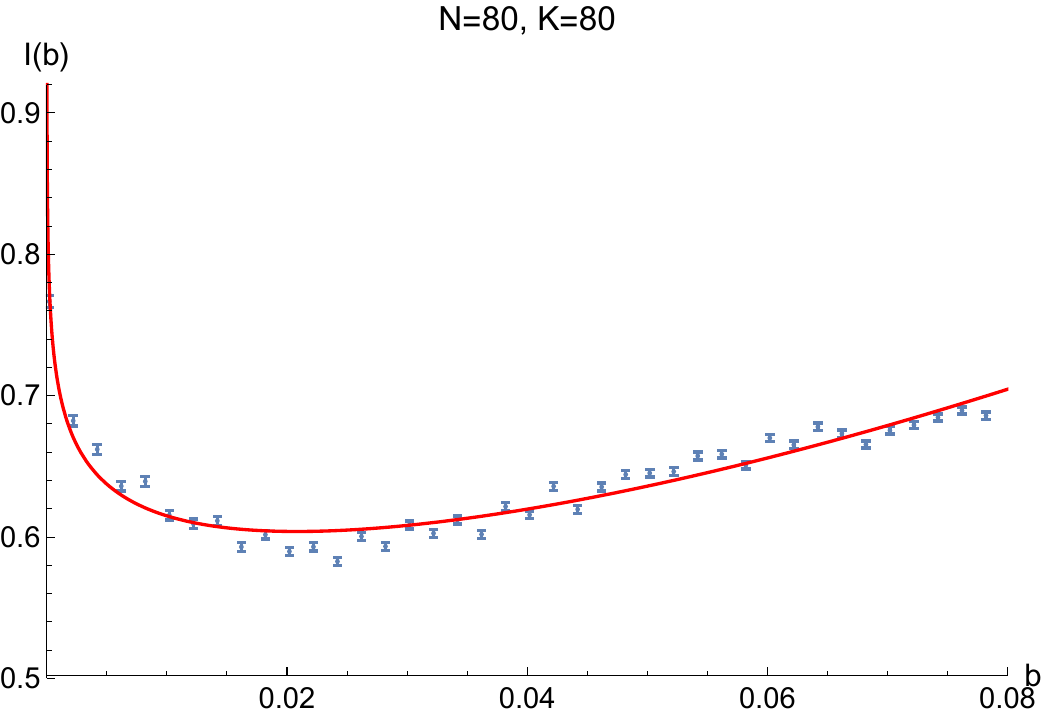}
	\includegraphics[width=0.45\textwidth]{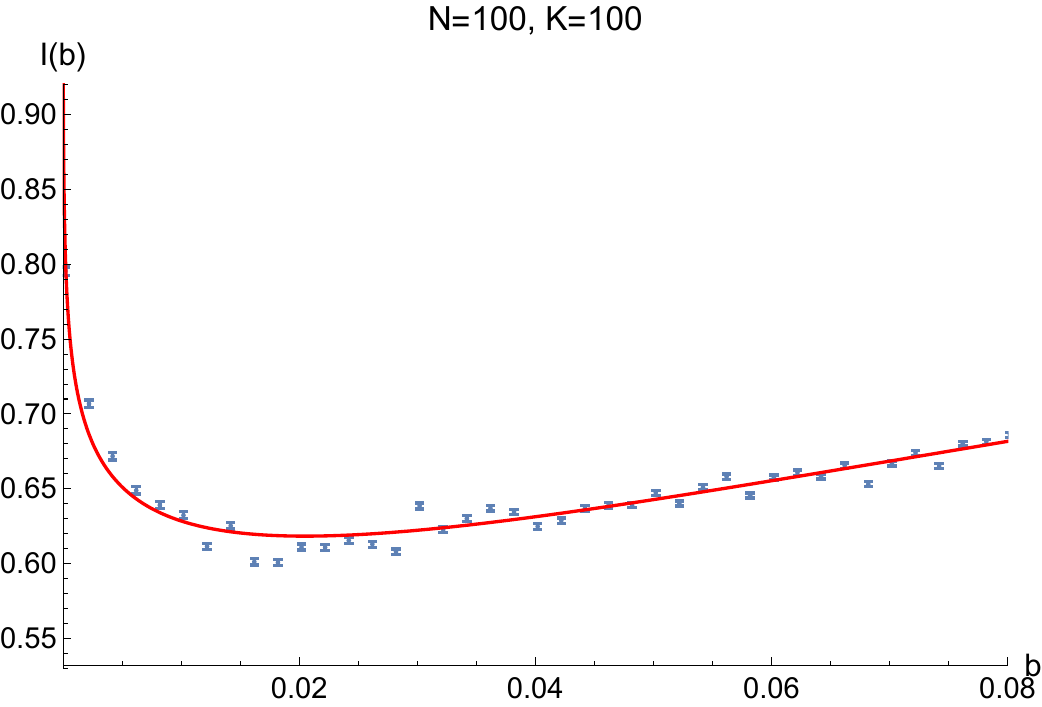}
	\caption{In blue, the averaged pairwise mutual information (measured in nats) for values of $b$ in $0.002$ intervals between $0.0002$ and $0.0782$, for different values of $N=K$. The points have error bars representing the standard error of the mean of the results from all the experiments. The red line is a result of a fit.}
	\label{fig:apmir1}
\end{figure}

\begin{figure}[htbp] 
	\includegraphics[width=0.55\textwidth]{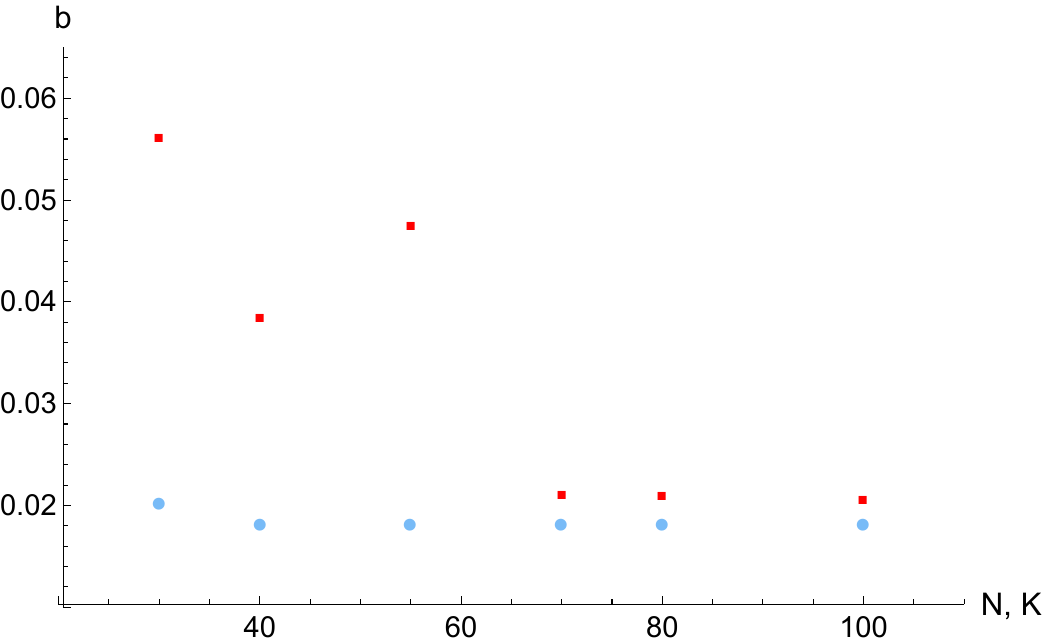}
	\caption{The blue disks: values of $b$ related to GUE type statistics for different $N=K$. The red squares: the associated values of $b$ for which the fit to the experimental APMI has a minimum.}
	\label{fig:br1}
\end{figure}

In figure \ref{fig:apmir1} we see the results of the experiment. This time, in each case, a function of the form 
\[f(b)=\alpha_0+\alpha_1 b+\alpha_2 b^2 +\alpha_3{\rm ln}(b)\]
 is fitted to the values of APMI. $\alpha_i$'s are the parameters of the fit. Now, the minimal values of APMI with respect to $b$ are found based on the fit of this function. We denote them by $b_{min}$ and plot them as red squares in figure \ref{fig:br1}. The blue disks, represent the values of $b$ associated with the best fit of the experimental nearest neighbor distribution to the GUE Wigner surmise. Let us call them $b_{GUE}$. For the random matrix statistics to emerge in the model, there needs to be a proper balance between the logarithmic repulsion term and the harmonic potential term of the utility function. When we keep the ratio of the number of buses and the number of stops constant, than the value of $b_{GUE}$ doesn't change, because all the variables in the utility function are the same - in other words, optimizing his next arrival time, the bus driver doesn't need to take into account the total number of buses, just their density. Hence, the associated $b_{GUE}$ values are roughly constant with respect to $N$, which can be seen on the plot. The values of $b_{min}$, converge to the value of $b_{GUE}$. We thus conclude, that, in the case of $N=K$, we can relay on the minimization rule for the pairwise mutual information when a sufficient number of buses participate in the process.  

\begin{figure}[htbp] 
	\includegraphics[width=0.45\textwidth]{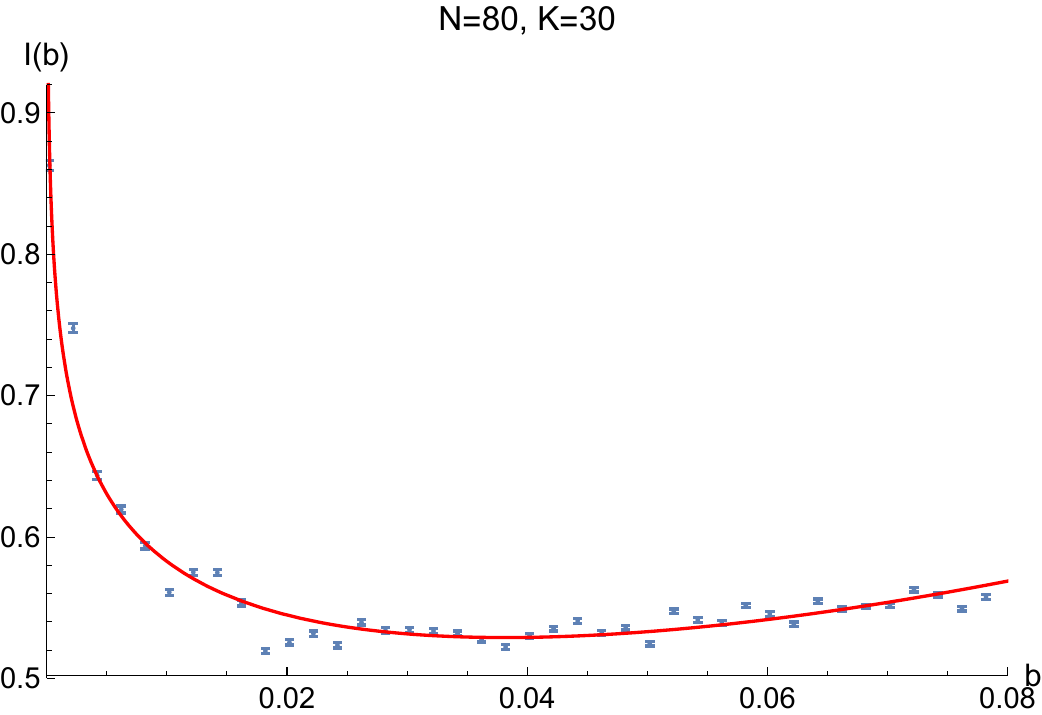}
	\includegraphics[width=0.45\textwidth]{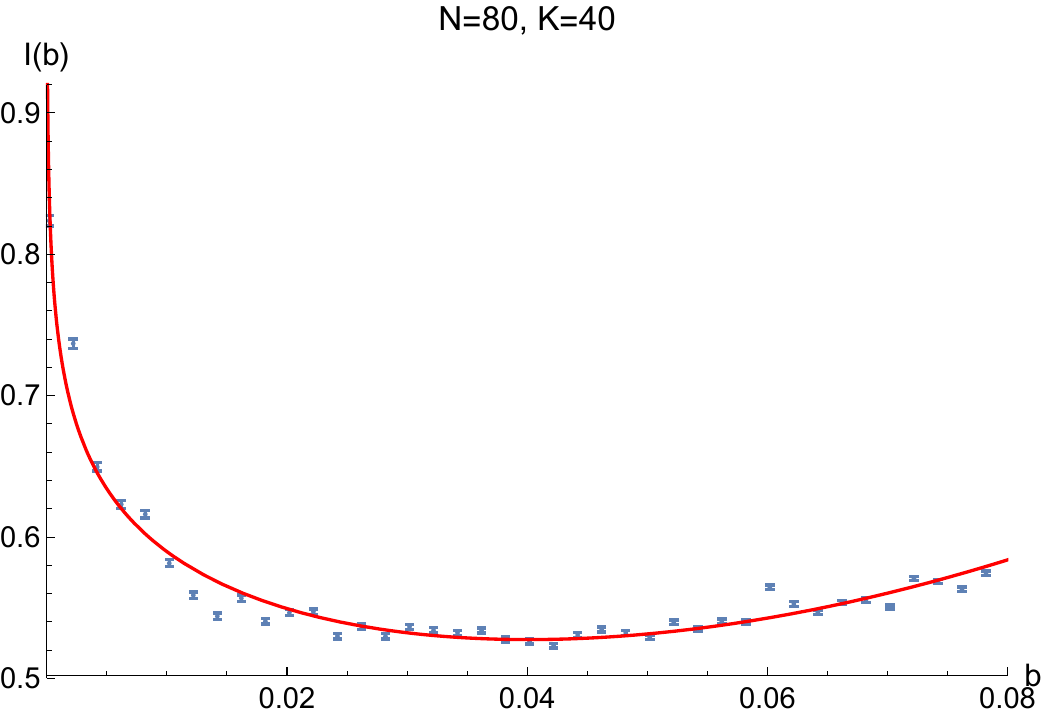}

	\includegraphics[width=0.45\textwidth]{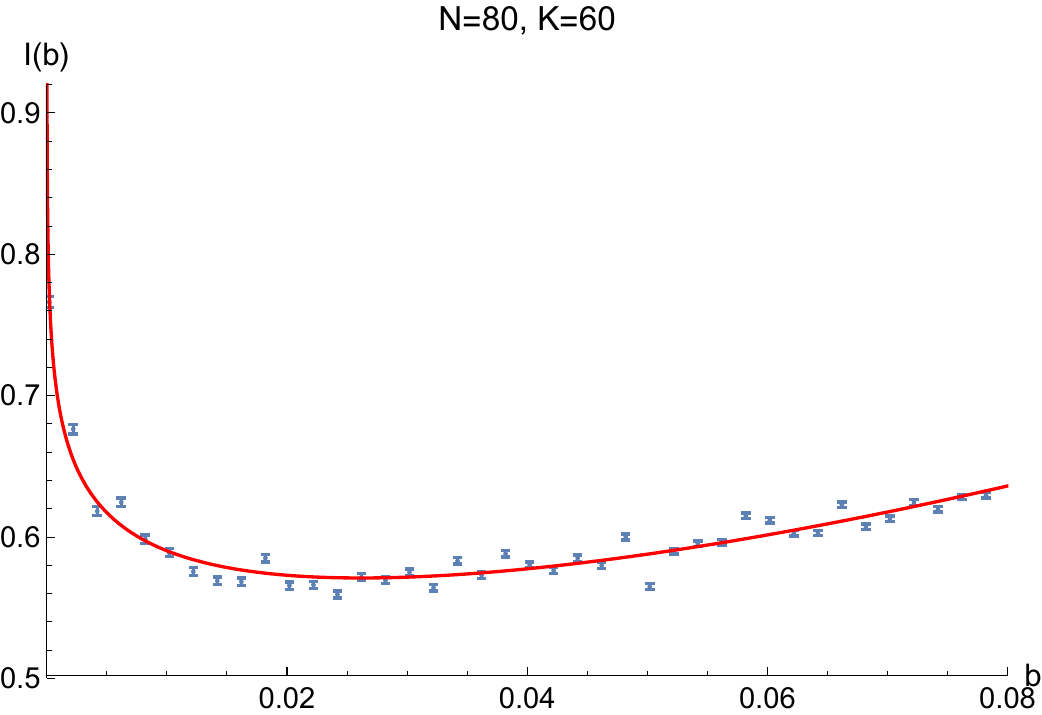}
	\includegraphics[width=0.45\textwidth]{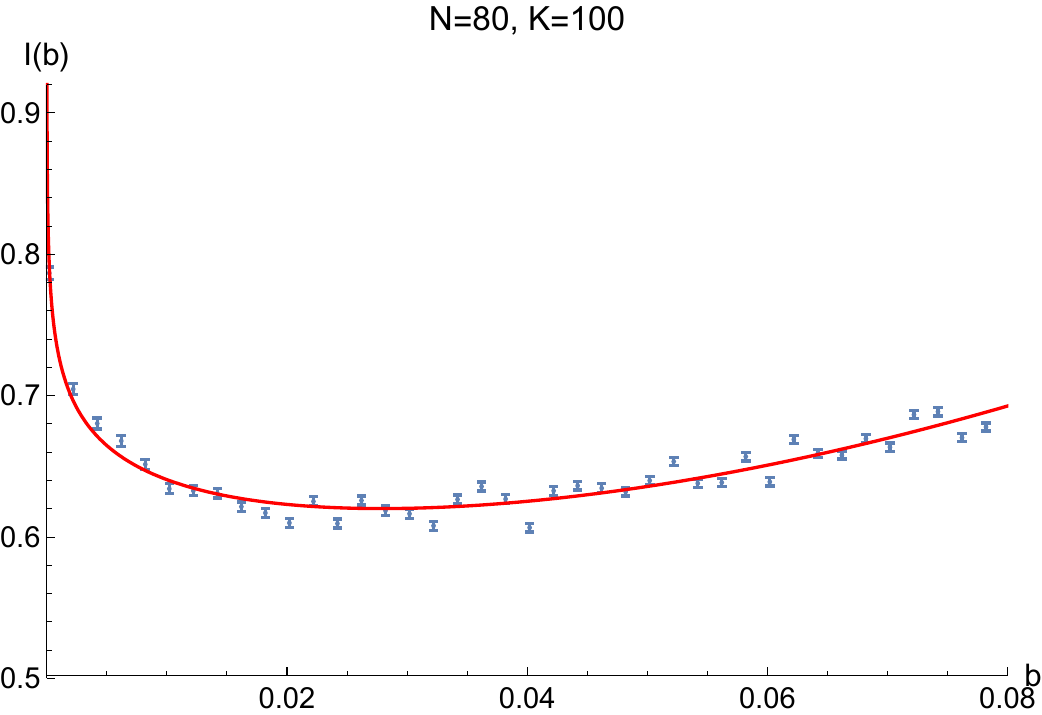}

	\includegraphics[width=0.45\textwidth]{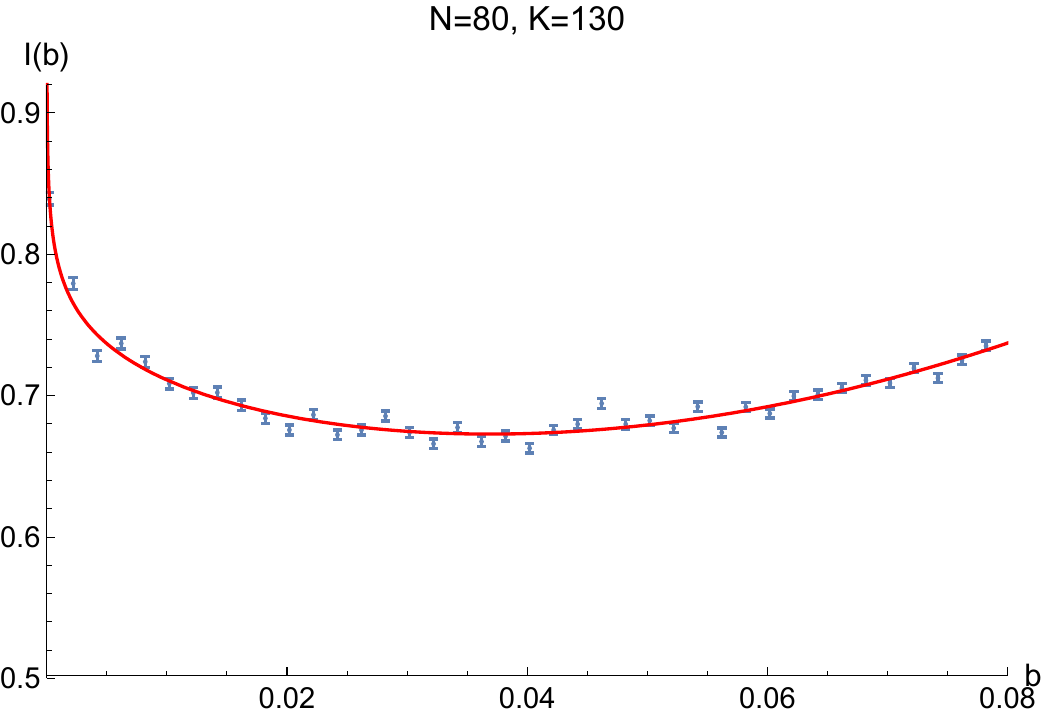}
	\includegraphics[width=0.45\textwidth]{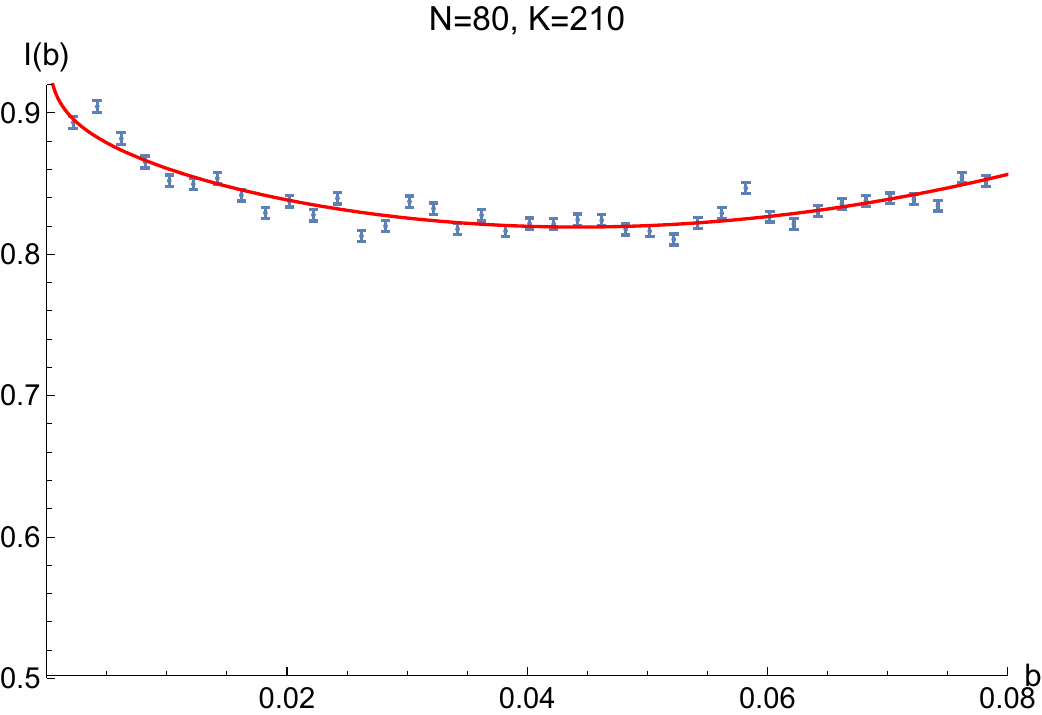}
	
	\includegraphics[width=0.45\textwidth]{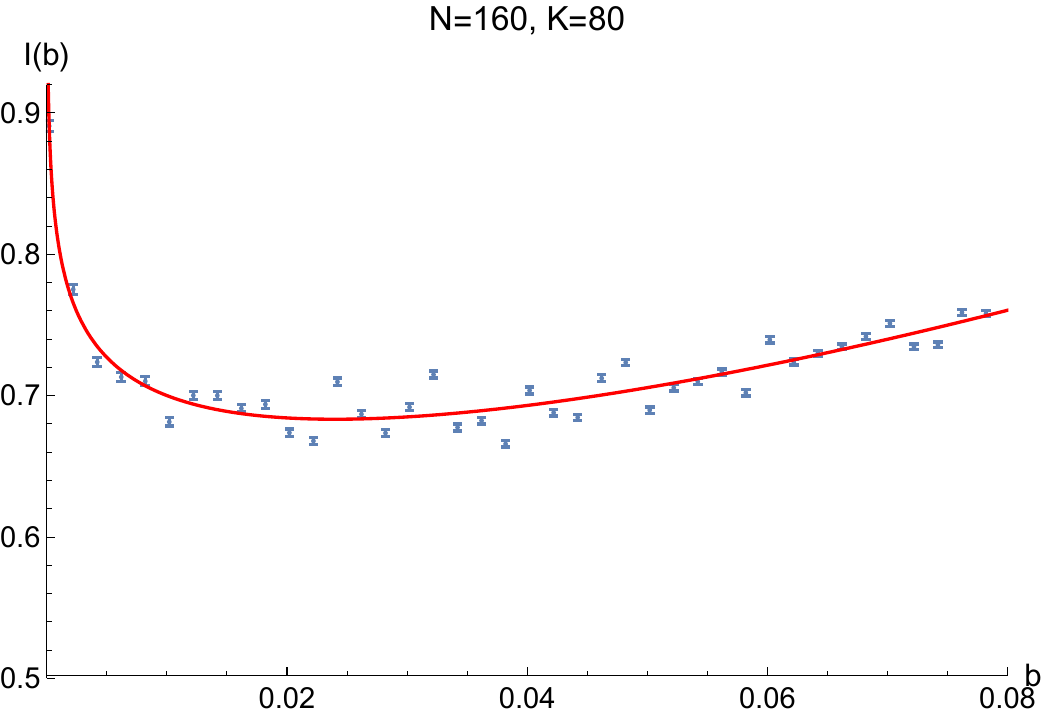}
	\caption{In blue, the averaged pairwise mutual information (measured in nats) for values of $b$ in $0.002$ intervals between $0.0002$ and $0.0782$, for different values of $N$ and $K$. The points have error bars representing the standard error of the mean of the results from all the experiments. The red line is a result of a fit.}
	\label{fig:apmichr}
\end{figure}

\begin{figure}[htbp] 
	\includegraphics[width=0.55\textwidth]{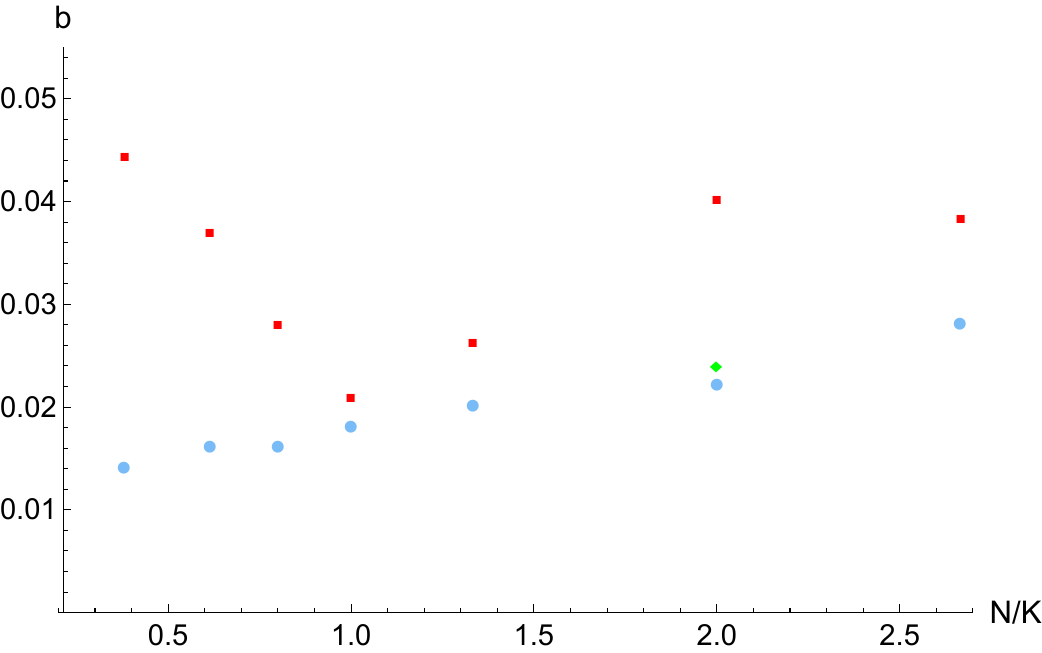}
	\caption{The blue disks: values of $b$ related to GUE type statistics for $N=80$ and different values $K$. The red squares: the associated values of $b$ for which the fit to the experimental APMI has a minimum. The green diamond: the value of $b$ for which the fit to the experimental APMI has a minimum in the case of the $N=160$, $K=80$ simulation.}
	\label{fig:bchr}
\end{figure}

Now we turn to the simulations for different $K$ and $N$. The results are shown in figure \ref{fig:apmichr}. We showcase the comparison of $b_{GUE}$ and $b_{min}$ in figure \ref{fig:bchr}. For bus densities (defined by $\frac NK$) outside of the vicinity of one, the APMI minimization rule `picks' $b$ values associated with statistics intermediate between GUE and Gaussian. This effect is more pronounced for $\frac NK<1$. Note, this is where the values of $N$ and $K$ used are equal or larger than $80$. In the case of $\frac NK>1$, as $N=80$, $K$ is smaller then $80$ and due to the conclusions of the previous paragraph, we decide to perform one more simulation with $N=160$ and $K=80$, doubling both numbers with respect to the associated simulation with $\frac NK=2$. The results are depicted in the last plot of figure \ref{fig:apmichr} and as the green (diamond shaped) point in figure \ref{fig:bchr}. The effect of increasing the number of buses and bus stops is the same as depicted in figure \ref{fig:br1}. Thus we conclude that to obtain decisive results on the behavior of APMI for the densities higher then one, even bigger simulations would need to be performed.

\onecolumngrid

\begin{thebibliography}{9}   


\bibitem{KSbuses}
M. \v{K}rbalek and P. \v{S}eba,
"The statistical properties of the city transport in Cuernavaca
(Mexico) and random matrix ensembles",
{\it J. Phys. A} {\bf 33} (2000) 229234.

\bibitem{RMTapli}
F. Dyson, foreword to {\it The Oxford Handbook of Random Matrix Theory} edit by G. Akemann, J. Baik, P. Di Francesco (Oxford University Press, 2011).

\bibitem{cars}
A. Y. Abul-Magd
"Modeling highway-traffic headway distributions using superstatistics",
{\it Phys. Rev. E} {\bf 76} (2007) 057101.

\bibitem{nycmetro}
A. Jagannath and T. Trogdon,
"Random matrices and the New York City subway system",
{\it Phys. Rev. E(R)}  {\bf 96} (2017) 030101.



\bibitem{BBDS}
J. Baik, A. Borodin, P. Deift and T. Suidan,
"A model for the bus system in Cuernavaca (Mexico)"
{\it J. Phys. A} {\bf 39} (2006) 8965.

\bibitem{KScella}
M. Krb\'alek and P. \v{S}eba,
"Headway statistics of public transport in Mexican cities",
{\it J. Phys. A} {\bf 36}(1) (2002) L7.


\bibitem{KHdue}
M. Krb\'alek and T. Hobza,
"Inner structure of vehicular ensembles and random matrix theory",
{\it J. Phys. A} {\bf 380}(1) (2016) 1839.


\bibitem{MIMO}
N. Bonneau, M. Debbah, E. Altman and A Hjorungnes,
"Non-atomic Games for Multi-User Systems",
{\it IEEE Journal on Selected Areas in Communications} {\bf 26} (2008) 1047.




\bibitem{NetworkinGames}
E. Altmana, T. Boulognea, R. El-Azouzia, T. Jim\'{e}nezb and L. Wynterc,
''A survey on networking games in telecommunications'',
{\it Computers and Operations Research} {\bf 33} (2006) 286 .


\bibitem{Wardrop}
J. G. Wardrop,
''Some theoretical aspects of road traffic research communication networks''
{\it  Proceedings of the Institution of Civil Engineers, Part 2} {\bf 1} (1952) 325.

\bibitem{Mehta} 
M. L. Mehta, \textit{Random Matrices} (Academic Press, New York, 1991), 2nd ed.     


\bibitem{SHte}
T. Schreiber, 
"Measuring Information Transfer", 
{\it Phys. Rev. Lett.} {\bf 85} (2000) 461.


\bibitem{JBCtecrit}
R. G. James, N. Barnett and J. P. Crutchfield,
"Information Flows? A Critique of Transfer Entropies",
{\it Phys. Rev. Lett.} {\bf 116} (2016) 238701.


\bibitem{Ising}
L. Barnett, J. T. Lizier, M. Harre, A. K. Seth and T. Bossomaier,
"Information Flow in a Kinetic Ising Model Peaks in the Disordered Phase",
{\it Phys. Rev. Lett.} {\bf 111} (2013) 177203.


\bibitem{Kraskov}
A. Kraskov, H. Stogbauer and P. Grassberger, 
"Estimating mutual information", 
{\it Phys. Rev. E} {\bf 69} (2004) 066138.
                
                
                                                                 
\bibitem{JIDT} 
J. T. Lizier, 
"JIDT: An information-theoretic toolkit for studying the dynamics of complex systems", 
{\it Frontiers in Robotics and AI} {\bf 1} (2014) 11.


\bibitem{DE}
I. Dumitriu and A. Edelman.
Matrix models for beta ensembles.
{\it J. Math. Phys.} {\bf 43} (2002) 5830. 

\bibitem{CMD}
G. Le Ca\"{e}r, C. Male and R. Delannay.
Nearest-neigbor spacing distributions of the $\beta$-Hermite ensemble of random matrices. 
{\it Physica A: Stat. Mech. and its Appl.}
{\bf 383} (2007) 190.

\bibitem{BEY}
Paul Bourgade, L\'{a}szl\'{o} Erd\H{o}s, and  Horng-Tzer Yau.
Universality of General $\beta$-Ensembles.
{\it Duke Math. J.} {\bf 163} (2014) 1127.


\end{thebibliography}
\end{document}